\documentclass[pra,twocolumn,showpacs]{revtex4}
\usepackage{amsmath,amssymb,mathrsfs}
\usepackage[applemac]{inputenc}
\usepackage{psfrag}
\usepackage{graphicx}
\usepackage{graphics}
\usepackage{epsfig}
\usepackage{bm}
\usepackage{color}
\usepackage{verbatim,color,ulem}

\newcommand{\beq}{\begin{equation}}
\newcommand{\eeq}{\end{equation}}
\newcommand{\beqa}{\begin{eqnarray}}
\newcommand{\eeqa}{\end{eqnarray}}

\begin{document}

\title{Quantum bright solitons in a quasi-one-dimensional optical lattice}
\author{Luca Barbiero$^{1}$ and Luca Salasnich$^{1,2}$}
\affiliation{$^{1}$Dipartimento di Fisica e Astronomia ``Galileo Galilei'' and 
CNISM, Università di Padova, Via Marzolo 8, 35131 Padova, Italy \\
$^{2}$Istituto Nazionale di Ottica (INO) del Consiglio Nazionale 
delle Ricerche (CNR), Sezione di Sesto Fiorentino, 
Via Nello Carrara, 1 - 50019 Sesto Fiorentino, Italy}

\date{\today}

\begin{abstract}
We study a quasi-one-dimensional attractive Bose gas confined 
in an optical lattice with a super-imposed harmonic potential 
by analyzing the one-dimensional Bose-Hubbard Hamiltonian of the system. 
Starting from the three-dimensional many-body quantum Hamiltonian 
we derive strong inequalities involving the transverse degrees of freedom
under which the one-dimensional Bose-Hubbard Hamiltonian can be safely used. 
In order to have a reliable description of the one-dimensional 
ground-state, that we call quantum bright soliton, we use the 
Density-Matrix-Renormalization-Group (DMRG) technique. 
By comparing DMRG results with mean-field (MF) ones 
we find that beyond-mean-field effects become relevant 
by increasing the attraction between bosons 
or by decreasing the frequency of the harmonic confinement. 
In particular we find that, contrary to the MF 
predictions based on the discrete nonlinear Schr\"odinger equation, 
average density profiles of quantum bright 
solitons are not shape invariant. 
We also use the time-evolving-block-decimation (TEBD)
method to investigate dynamical properties 
of bright solitons when the frequency of the harmonic 
potential is suddenly increased. This quantum quench induces 
a breathing mode whose period crucially depends on the final 
strength of the super-imposed harmonic confinement. 
\end{abstract}


\maketitle

\section{Introduction}

Ultracold bosonic gases in reduced dimensionality are an ideal platform 
for probing many-body phenomena where quantum fluctuations 
play a fundamental role~\cite{giamarchi,cazalilla2011}. 
In particular, the use of optical lattices has allowed the experimental 
realization  \cite{bloch} of the well-known 
Bose-Hubbard Hamiltonian \cite{fisher} with dilute and 
ultracold alkali-metal atoms. 
This achievement has been of tremendous impact on several 
communities \cite{book-lattice} since it is one of the 
first experimental realization of 
a model presenting a pure quantum phase transition, namely the metal-Mott 
insulator transition. At the same time new experimental techniques, like 
in-situ imaging \cite{insitu}, are now available 
to detect many-body correlations and density profiles. 
Furthermore, these techniques offer the possibility to observe 
intriguing many-body effects in regimes which are far from equilibrium.  
In this contest the relaxation dynamics regimes \cite{relaxation} 
and light-cone-like effects \cite{kollath} in a one-dimensional (1D) Bose 
gas loaded on a optical lattice have been recently observed. 

The 1D Bose-Hubbard Hamiltonian, which accurately 
describes dilute and ultracold atoms in a strictly 1D optical lattice, 
is usually analyzed in the case of repulsive interaction strength
which corresponds to a positive inter-atomic s-wave scattering length 
\cite{bloch_review}. Indeed, a negative s-wave 
scattering length implies an attractive interaction strength 
which may bring to the collapse \cite{brand,sala-dnpse}
due to the shrink of the transverse width of a realistic quasi-1D bosonic 
cloud. Moreover in certain regimes of interaction 
the quasi-1D mean-field (MF) theory 
predicts the existence of meta-stable configurations \cite{kevre}  
which are usually called discrete bright solitons. 
We remark that continuous bright solitons have been observed 
in various experiments \cite{exp-solo1,exp-solo2,exp-solo3,exp-solo4} 
involving  attractive bosons of $^7$Li and $^{85}$Rb vapors. Instead, 
discrete (gap) bright solitons in quasi-1D optical lattices 
have been observed \cite{exp-gap} 
only with repulsive bosons made of $^{87}$Rb atoms. 

In this paper we first derive an effective 1D Bose-Hubbard Hamiltonian 
which takes into account the transverse width of the 3D atomic cloud. 
In this way we determine a strong inequality under 
which the effective 1D Bose-Hubbard Hamiltonian reduces 
to the familiar one and the collapse of discrete bright solitons 
is fully avoided. We then work in this strictly 1D regime 
analyzing the 1D Bose-Hubbard Hamiltonian by using 
the Density-Matrix-Renormalization-Group (DMRG) technique \cite{white}. 
We evaluate density profiles and quantum fluctuations 
finding that, for a fixed number of atoms, there are regimes where 
the MF results (obtained with a discrete nonlinear 
Schr\"odinger equation) strongly differ from the DMRG ones. 
Finally, we impose a quantum quench to the discrete bright solitons 
by suddenly increasing the frequency of the harmonic potential. 
By using the  time-evolving-block-decimation (TEBD) 
method \cite{vidal} we find that 
this quantum quench induces a breathing oscillation 
in the bosonic cloud. Also in this dynamical case we find that 
the MF predictions are not reliable when the on-site 
attractive energy is large. 

\section{The model}

We consider a dilute and ultracold gas of bosonic atoms confined 
in the plane $(x,y)$ by the transverse harmonic potential
\beq
U(x,y) = {m\over 2} \omega_{\bot}^2
\left(x^2 + y^2 \right) \; . 
\eeq
In addition, we suppose that the axial potential is the 
combination of periodic and harmonic potentials, i.e. 
\beq
V(z) = V_0 \cos^2{(2 k_0 z)} + {1\over 2} \omega_z^2 z^2 \; .  
\eeq
This potential models the {quasi-1D optical lattice} 
produced in experiments with 
Bose-Einstein condensates by using counter-propagating 
laser beams \cite{morsch}. 
We choose $\lambda=\omega_z/\omega_{\bot} \ll 1$ which implies 
a weak axial harmonic confinement. 
The characteristic harmonic length of transverse confinement is given by 
$a_{\bot}=\sqrt{\hbar/(m\omega_{\bot})}$ 
and, for simplicity, we choose $a_{\bot}$ and $\hbar \omega_{\bot }$ 
as length unit and energy unit respectively. In the rest of the paper 
we use scaled variables. 

We assume that the system is well described by the 
quantum-field-theory Hamiltonian (in scaled units) 
\beqa
H = \int d^3{\bf r} \ { \psi}^+({\bf r}) 
\Big[ -{\frac{1}{2}}\nabla^{2}
+ U(x,y) + V(z) 
\nonumber
\\
+ \pi g \ { \psi}^+({\bf r})  { \psi}({\bf r}) 
\Big] { \psi}(\mathbf{r}) \; , 
\label{3dgpe}
\eeqa
where ${ \psi}(\mathbf{r})$ is the bosonic field operator 
and $g= 2a_s/a_{\bot}$ with $a_s$ the s-wave scattering length of the
inter-atomic potential \cite{book-bose}.
 
\subsection{Discretization}

We perform a discretization of the 3D Hamiltonian along the $z$ 
axis due to the presence on the periodic potential. 
In particular we use the decomposition \cite{book-lattice}
\beq
\psi(\mathbf{r}) = \sum_i { \phi}_i(x,y) \ w_i(z) 
\eeq
where $w_i(z)$ is the {Wannier function} maximally localized 
at the $i$-th minimum of the axial periodic potential. 
In this paper we consider the case of an even number $L$ of sites 
$z_i=(2i-L-1)\pi/(4k_0)$ with $i=1,2,...,L$. 

This tight-binding ansatz is reliable in the case of a deep optical 
lattice \cite{trombettoni} \cite{zoller}. 
To further simplify the problem we set (field-theory 
extension of the mean-field approach developed in \cite{sala-dnpse})  
\beq
\phi_i(x,y) |GS\rangle = 
{1\over \pi^{1/2} \sigma_i}
\exp{\left[ - \left( {x^2+y^2\over 2\sigma_i^2} 
\right) \right] }\, { b}_i |GS\rangle \; , 
\label{assume}
\eeq
where $|GS\rangle$ is the many-body ground state,  while 
$\sigma_i$ and ${ b}_i$ account respectively 
for the adimensional on-site transverse width (in units of $a_{\bot}$) 
and for the bosonic field operator. 
We insert this ansatz into Eq. (\ref{3dgpe}) and 
we easily obtain the effective 1D Bose-Hubbard Hamiltonian 
\beqa 
H &=& \sum_{i} \Big\{ 
\big[ {1\over 2} ( {1\over \sigma_i^2} +  
\sigma_i^2) + \epsilon_i \big] { n}_i 
\nonumber
\\
&-& J \, { b}_i^+ \left( { b}_{i+1}+{ b}_{i-1} \right) 
+ {1\over 2} {U \over \sigma_i^2} { n}_i ({ n}_i-1) \Big\} \; ,  
\label{e1}
\eeqa
where ${ n}_i={ b}_i^+{ b}_i$ is the on-site number operator, 
\beq 
\epsilon_i = \int w_{i}^*(z) \left[ -{1\over 2}
{\partial^2\over \partial z^2} 
+ V(z) \right] w_i(z) \ dz 
\eeq
is the on-site axial energy, which can be written as 
$\epsilon_i=V_S + V_T \ (2i-L-1)^2$ 
with $V_S$ the on-site energy due to the periodic potential and 
$V_T$ the strength (harmonic constant) 
of the super-imposed harmonic potential, while $J$ and $U$ 
are the familiar adimensional hopping (tunneling) energy and 
adimensional on-site energy which are 
experimentally tunable via $V_0$ and $a_s$ \cite{bloch_review}. 
$J$ and $U$ are given by 
\beqa
J &=& - \int w_{i+1}^*(z) \left[ -{1\over 2}
{\partial^2\over \partial z^2} 
+ V(z) \right] w_i(z) \ dz  \; , 
\\
U &=& g \int |w_i(z)|^4 \ dz  \; . 
\eeqa
Remember that now $V(z)$ is in units of $\hbar\omega_{\bot}$ and 
$z$ in units of $a_{\bot}$. 
Even if $J$ and $U$ actually depend on the site index $i$, the choice 
of considering low $V_T$ allows us to keep them constant. 

\subsection{Dimensional reduction}

Our Eq. (\ref{e1}) takes into account deviations with respect 
to the strictly 1D case 
due to the transverse width $\sigma_i$ of the bosonic field. 
We call the Hamiltonian (\ref{e1}) quasi-1D because it depends 
explicitly on a transverse width $\sigma_i$ which is not 
equal to the characteristic length $a_{\bot}$ of the transverse 
harmonic confinement.

This on-site transverse width $\sigma_i$ can be determined 
by averaging the Hamiltonian (\ref{e1}) over a many-body 
quantum state $|GS\rangle$ and minimizing the resulting energy function 
\beqa 
\langle GS|H|GS \rangle &=& \sum_{i} \Big\{ 
\big[ {1\over 2} ( {1\over \sigma_i^2} +  
\sigma_i^2) + \epsilon_i \big] \langle { n}_i \rangle 
\nonumber
\\
&+& {1\over 2} {U \over \sigma_i^2} \left( \langle { n}_i^2\rangle - 
\langle { n}_i\rangle \right)
\\
&-& J \,  \left( \langle { b}_i^+ { b}_{i+1} \rangle 
+ \langle { b}_i^+ { b}_{i-1} \rangle \right)  \Big\} 
\nonumber
\eeqa
with respect to  $\sigma_i$. Notice that the hopping term 
is independent on $\sigma_i$. In this way, by using the Hellmann-Feynman 
theorem, one immediately gets 
\beq
\sigma_i^4 = 1 + U {\langle { n}_i^2\rangle - \langle { n}_i\rangle 
\over \langle { n}_i \rangle }\; . 
\label{sig1} 
\eeq
Eqs. (\ref{e1}) and (\ref{sig1}) must be solved 
self-consistently to obtain the ground-state of the system. 
Clearly, if $U<0$ the transverse width $\sigma_i$ is smaller than 
one (i.e. $\sigma_i <a_{\bot}$ in dimensional units) and 
the collapse happens when $\sigma_i$ goes to zero 
\cite{sala-dnpse}. At the critical strength $U_c$ of the collapse 
all particles are accumulated in the two central sites 
($i=L/2$ and $i=L/2+1$) around the minimum of the harmonic potential 
and consequently $U_c \simeq -2/N$ 
(i.e. $U_c/(\hbar \omega_{\bot}) \simeq -2/N$ in dimensional units). 

We stress that, from Eq. (\ref{sig1}), the system is strictly 
1D only if the following strong inequality 
\beq 
U  {\langle { n}_i^2\rangle - \langle { n}_i\rangle 
\over \langle { n}_i \rangle }  \ll 1  
\label{condition}
\eeq 
is satisfied for any $i$, such that $\sigma_i=1$ 
(i.e. $\sigma_i=a_{\bot}$ in dimensional units). Under 
the condition (\ref{condition}) the problem of collapse 
is fully avoided. 

\begin{figure}[t]
\centerline{\epsfig{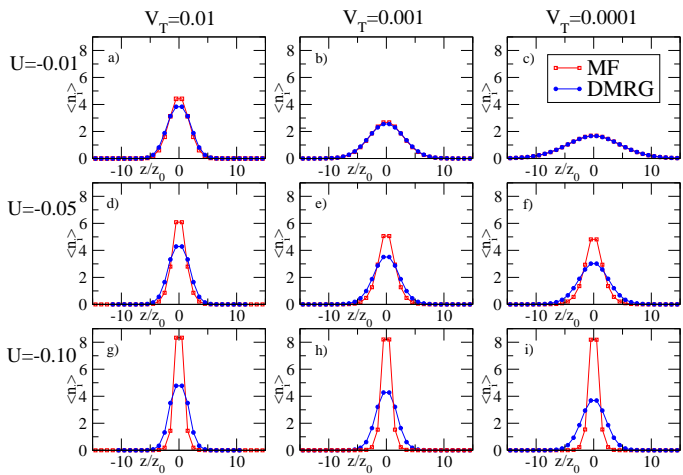}}
\caption{(Color online). 
MF (squares) and DMRG (circles) density profiles 
$\langle n_i \rangle$ of the bright soliton 
with $J=0.5$, $L=80$ and $N=20$. In the horizontal axis 
there is the scaled axial coordinate $z/z_0$, with 
$z_0=\pi/(4k_0)$. The results of each panel are obtained with 
different values of harmonic strength $V_T$ (columns) 
and interaction strength $U$ (rows).}
\label{fig1}
\end{figure}

\section{Numerical results}

In the remaining part of the paper we shall work in this strictly 
1D regime where the effective Hamiltonian of 
Eq. (\ref{e1}) becomes (neglecting 
the irrelevant constant transverse energy) 
\beq 
H =-J \sum_i b_i^\dagger (b_{i+1} + b_{i-1})  +\frac{U}{2}
\sum_in_i(n_i-1)+ \sum_i \epsilon_i \ n_i 
\label{ee1}
\eeq
which is the familiar 1D Bose-Hubbard model \cite{fisher}. 
We call the Hamiltonian (\ref{ee1}) strictly-1D because 
the transverse width $\sigma_i$ is equal to the characteristic 
length $a_{\bot}$ of transverse harmonic confinement. 

\subsection{Glauber coherent state and DNLSE}

As already mentioned, in a 1D configuration quantum fluctuations, 
which are actually neglected in mean-field approaches, 
can play a relevant role. Thus, in our 1D problem it is relevant 
to compare MF predictions with DMRG ones 
in order to observe in which regimes MF can give accurate 
and reliable results. 
In particular, we use a MF approach based on Glauber coherent state
\begin{equation}
|GCS\rangle = |\beta_1 \rangle \otimes ... \otimes |\beta_L\rangle
\end{equation} 
where $|\beta_i\rangle$ is, by definition, such that 
$b_i |\beta_i\rangle =\beta_i |\beta_j\rangle$ \cite{book-luca}. 
By minimizing the energy 
$\langle GCS|H|GCS\rangle$ with respect to $\beta_i$, one finds  
that the complex numbers $\beta_i$ satisfy the 1D discrete  
nonlinear Schr\"odinger equation (DNLSE) 
\beq 
\mu \, \beta_i = \epsilon_i \, \beta_i 
- J \, \left( \beta_{i+1} +\beta_{i-1} \right) + U |\beta_i|^2 \beta_i \; ,  
\label{dgpe}
\eeq
where $\mu$ is the chemical potential of the system fixed 
by the total number of atoms: 
$N=\sum_i |\beta_i|^2=\sum_i \langle GCS|{ n}_i|GCS\rangle$. 
By solving Eq. (\ref{dgpe}) with Crank-Nicolson predictor-corrector 
algorithm with imaginary time 
\cite{sala-numerics} it is possible to show that in 
the attractive case ($U<0$) discrete bright solitons exist \cite{sala-dnpse}. 

\begin{figure}[t]
\centerline{\epsfig{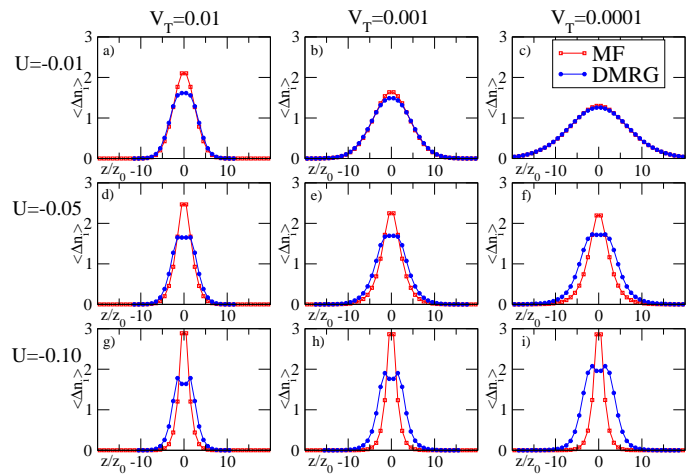}}
\caption{(Color online).  
MF (squares) and DMRG (circles) quantum fluctuations 
$\Delta n_i=\sqrt{\langle n^2_i\rangle-\langle n_i\rangle^2}$ 
of the bright soliton with $J=0.5$, $L=80$, 
and $N=20$.  In the horizontal axis 
there is the scaled axial coordinate $z/z_0$, with 
$z_0=\pi/(4k_0)$. The results of each panel are obtained with 
different values of harmonic strength $V_T$ (columns) 
and interaction strength $U$ (rows).}
\label{fig2}
\end{figure}

On general physical grounds one expects that 
the MF results obtained from the DNLSE of Eq. (\ref{dgpe}) 
are fully reliable only when $U\to 0$ and $N\to \infty$ 
with $UN$ taken constant. Indeed the Glauber coherent state $|GCS\rangle$
is the exact ground state of the Bose-Hubbard Hamiltonian 
only if $U=0$ and $N\to \infty$. 
Notice that the exact ground state of the Bose-Hubbard 
Hamiltonian with $U=0$ and a finite number $N$ of bosons is the 
atomic coherent state $|ACS\rangle$ which reduces to the Glauber 
coherent state $|GCS\rangle$ only for $N\to \infty$ (see for instance 
\cite{atomic}). In practice, one expects that MF results are 
meaningful in the superfluid regime where there is a quasi-condensate, 
i.e. algebric decay of phase 
correlations \cite{giamarchi,book-lattice,book-bose}. Nevertheless, 
in general it is quite hard to determine this superfluid regime. 
For this reason, working with a 
small number $N$ of bosons it is important to compare the Glauber 
MF theory with a quasi-exact method. 

\subsection{DMRG approach}

DMRG is able to take into account the full quantum behavior 
of the system and it has already given strong evidences of solitonic waves 
in spin chains \cite{manmana} and in bosonic models with nearest 
neighbors interaction \cite{das}.  A crucial point in order to have accurate 
results by using DMRG is played by the size of the Hilbert space we set 
in our simulations. Clearly, for system sizes and densities 
comparable with the experimental ones, we can not investigate the collapse 
phase where all the bosons "collapse" in one site. Indeed it requires a 
size of the Hilbert space which is not approachable with our method. 
Anyway, as shown in Eq. (\ref{sig1}) this phase does 
not happen for sufficiently 
low density and on-site interaction $U$. For this reason and in order 
to fulfill Eq. (\ref{sig1}) we consider regimes which are sufficiently 
far from this scenario, more precisely we use a number $N=20$ of 
bosons in $L=80$ lattice 
sites and interactions $U\ge-0.1$. Nevertheless if we allow 
a too small number of bosons per site, namely if we consider 
a too small Hilbert space, even if we are far from the 
collapse, our results might be not reliable since the shape of density 
profile is modified by this cut off and not by physical reasons.
To treat this problem we consider a maximum number of bosons 
in each site $n_{max}=8$ and we checked that increasing this quantity 
does not significantly affect our results. Moreover we keep up to $512$ 
DMRG states and 6 fine size sweeps \cite{white} to have a truncation 
error lower than $10^{-10}$.

\subsection{Comparing DMRG with DNLSE}
 
In Fig. \ref{fig1} we compare the density 
profiles given by DMRG with the ones obtained by using 
the mean-field DNLSE for different strengths of the 
harmonic potential and interaction. For weak interactions $U$ the particles 
are substantially free and the shape of the cloud is given only by 
the harmonic strength $V_T$. Of course when the particles 
are strongly confined in the center of the system, 
as in panel $a)$ of Fig. \ref{fig1}, 
the interaction $U$ begins to play a role due to the relevant number of bosons 
lying in the two central sites. More precisely $U$ tries to drop 
quantum fluctuations induced by $J$ and it explains the small but 
significant discrepancies we find. 

When the interaction $U$ 
is sufficiently strong (panels $g),h),i)$ of Fig. \ref{fig1}) 
the MF results becomes insensitive to the super-imposed 
harmonic potential of strength $V_T$ 
since the shape of the cloud remains practically 
unchanged giving rise to self-localized profiles. Instead 
DMRG results do not show this self-localization.  
In fact quantum fluctuations try, in opposition to $U$, 
to maximally delocalize the bosonic cloud. 

In order to check if our interpretation is valid we plot 
in Fig. \ref{fig2} the expectation value of quantum fluctuations 
\begin{equation}
\Delta n_i=\sqrt{\langle n_i^2\rangle-\langle n_i\rangle^2} \; . 
\label{osc}
\end{equation}
For the Glauber coherent state $|GCS\rangle$ 
one has $\Delta n_i=\sqrt{n_i}$. We expect that $\Delta n_i$ 
of the DMRG ground-state $|GS\rangle$ 
can be quite different from the MF prediction. More precisely 
quantum fluctuations 
are enhanced by the kinetic term $J$ 
which tries to maximally spread the shape of the cloud. On the other hand 
the value of $\Delta n_i$ is minimized both by the strong on-site interaction 
(because the system gains energy having many particles in the same 
site) and by the strong trapping potential (which confines the bosons in 
the two central sites of the lattice where $V_T$ is weaker). 
This behavior is clear in Fig. \ref{fig2} where, for large $U$ and small $V_T$, 
$\Delta n_i$ presents strong deviations from MF behavior,  
whereas mean-field DNLSE and DMRG are in substantial agreement 
in the opposite regime. 

\begin{figure}[t]
\centerline{\epsfig{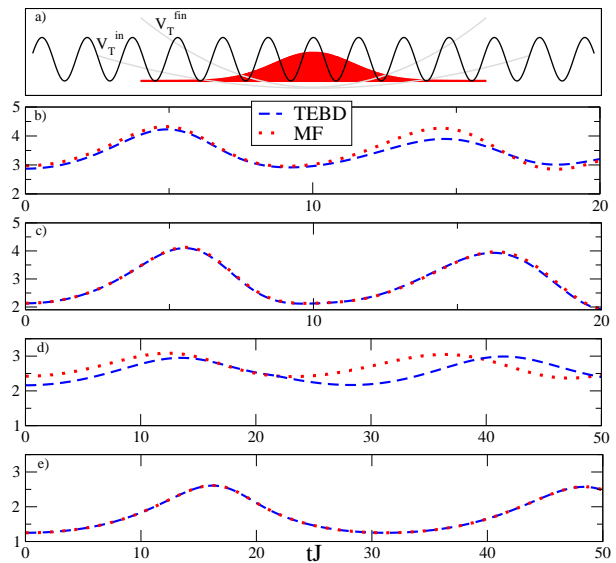}}
\caption{(Color online). $a)$ cartoon of the quench procedure. 
In the other panels: 
MF (dotted lines) and TEBD (dashed lines) 
central density $\langle n_{L/2} \rangle$ 
vs time $t$ with $J=0.5$, $L=40$, and $N=10$.  
Panels $b,c,$: quench from $V_T^{in}=0.01$ to 
$V_T^{fin}=0.05$ for respectively $U=-0.1, -0.01$. 
Panels: $d,e$ quench from $V_T^{in}=0.001$ to $V_T^{fin}=0.005$ for 
respectively $U=-0.1, -0.01$.}
\label{fig3}
\end{figure}
\subsection{Dynamical properties}
Another relevant aspect of bright solitons is given by its dynamical 
properties. In particular it is predicted by time-dependent 
DNLSE \cite{kevre} that discrete bright soliton can give rise 
to a breathing mode. To study the time evolution of the system 
we use the time-evolving-block-decimation (TEBD) algorithm \cite{vidal}, 
which is still a quasi-exact method 
recently used to study the appearance of a dark soliton \cite{sacha2} 
and its entanglement properties\cite{carr} \cite{carr2}. 
We compare TEBD results with time-dependent DNLSE ones, which  
are immediately obtained from Eq. (\ref{dgpe}) 
with the position $\mu \to i d/dt$. 
We determine the ground-state of the Bose system 
for a chosen value $V_T^{in}$ of transverse confinement 
and then we perform the time evolution 
with a larger value $V_T^{fin}$. In this way 
we mimic a sudden change in the strength $V_T$ of the 
super-imposed harmonic confinement (see panel a) of Fig. \ref{fig3}). 

In panels $b),c),d),e)$ of Fig. \ref{fig3} we report 
the density of atoms in the two 
central sites (where it takes the highest value since the effect of $V_T$ 
is weaker) as a function of time $t$. 
The panels show a periodic oscillation where 
the period $\tau$ of this breathing 
mode strongly grows by reducing the harmonic strength $V_T$. Moreover, 
$\tau$ is slightly enhanced by a smaller $|U|$. Remarkably, 
as in the static case, beyond-mean-field effects 
become relevant for a strong $|U|$ and they are instead less evident for 
high values of $V_T$. Indeed, in Fig. \ref{fig3} the relative difference 
between TEBD and MF in the period $\tau$ 
is below $1\%$ in panels $c)$ and $e)$, while it is around $8\%$ in 
panel $b)$ and around $37\%$ in panel $d)$.  

\section{Conclusions}

In this paper we have obtained 
a strong inequality, Eq. (\ref{condition}), under which the 3D system 
is reduced to a strictly-1D one and the collapse is fully avoided. Moreover, 
we have compared MF theory with the DMRG looking for 
beyond-mean-field effects in the effective 1D system of bosons in a lattice. 
From our results we conclude that the self-localized 
discrete bright solitons obtained by 
the MF nonlinear Schr\"odinger equation 
are not found with the DMRG results (quantum bright solitons). 
In other words, we have found that with a small number $N$ of bosons 
the average of the quantum density profile, that is experimentally obtained 
with repeated measures of the atomic cloud, is not shape invariant.  
This remarkable effect can be explained by considering a quantum 
bright soliton as a MF bright soliton with a center of mass 
\cite{calogero,castin} that is randomly distributed 
due to quantum fluctuations, 
which are suppressed only for large values of $N$ \cite{explain}. 
This is the same kind of reasoning adopted some years ago 
to explain the distributed vorticity of superfluid 
liquid $^4$He \cite{luciano}, and, more recently,
the Anderson localization of particles 
in one dimensional system \cite{sacha1} and
the filling of a dark soliton \cite{sacha2}. 
For the sake of completeness, 
we have also analyzed the breathing mode of discrete bright solitons 
after a sudden quench finding that also in the dynamics 
beyond mean-field effects become relevant for a strong interaction strength 
$U$ and for a small harmonic constant $V_T$ \cite{newcit}. 
Our paper gives strong evidence on the limits of MF theory 
in the study of bright solitons, suggesting that DMRG calculations 
must be used to simulate and analyze them. \\

The authors acknowledge for partial support 
Universit\`a di Padova (grant No. CPDA118083), 
Cariparo Foundation (Eccellenza grant 11/12), 
and MIUR (PRIN grant No. 2010LLKJBX). 
The authors thank L.D. Carr, B. Malomed, 
S. Manmana, A. Parola, V. Penna, and F. Toigo for fruitful 
discussions,  and S. Saha for useful suggestions.


\begin{thebibliography}{99}
\bibitem{explain} The axial width $\eta$ of the quantum 
bright soliton can be estimated as 
$\eta_{ES}=(\eta_{MF}^2+\eta_{QF}^2)^{1/2}$, where 
$\eta_{MF}$ is the width of the MF bright soliton and 
$\eta_{QF}= (a_s/a_{\bot}) a_z/N^{1/4}$ is the 
contribution due to quantum fluctuations of the center of mass, with 
$a_z$ the characteristic length of the super-imposed 
harmonic confinement, $a_s$ the s-wave 
scattering length, $a_{\bot}$ the characteristic length 
of the transverse harmonic confinement, and $N$ the number of atoms. 
We have verified that the estimated width $\eta_{ES}$ 
gives a rough semi-quantitative description of the axial width $\eta$ 
calculated with the DMRG approach.
\bibitem{newcit} The spreading of quantum bright solitons over time 
was investigated without harmonic trapping both in the 
continuum \cite{newcit1} and on the lattice \cite{newcit2}.
\end{thebibliography}
\end{document}